\documentclass[final,5p,times,twocolumn]{elsarticle}
\usepackage[
colorlinks=true,
linkcolor=blue,
urlcolor=blue,
citecolor=blue,
pdftex,
]{hyperref}
\usepackage{amsmath,graphicx,tikz}
\DeclareMathOperator*{\argmin}{argmin}

\newcommand*{\argminl}{\argmin\limits}

\usepackage{dsfont}

\usepackage{subfigure}
\usepackage{multirow}
\usepackage{array}
\usepackage{colortbl}
\usepackage{tabularx}
\usepackage{enumerate}

\usepackage{amsmath}
\usepackage{amssymb}

\usepackage{booktabs}
\usepackage{rotating}
\graphicspath{{figures/}}
\usepackage{threeparttable}
\usepackage[utf8]{inputenc} 

\usepackage{todonotes}

\usepackage{algorithm}
\usepackage{algpseudocode}

\usepackage{enumitem}

\journal{Pre-print}

\begin{document}

\begin{frontmatter}



\title{Left/Right Hand Segmentation in Egocentric Videos}


\author{
    \begin{tabular*}{0.9\textwidth}{c
 @{\extracolsep{\fill}} c @{\extracolsep{\fill}} c} Alejandro Betancourt$^{1,2}$& Pietro Morerio$^{1}$ & Emilia Barakova$^{2}$ \\ 
 \texttt{ \small a.betancourt@tue.nl} & \texttt{ \small pmorerio@ginevra.dibe.unige.it} & \texttt{\small e.i.barakova@tue.nl} \\
 \\
 Lucio Marcenaro$^{1}$& Matthias Rauterberg$^{2}$ & Carlo Regazzoni$^{1}$ \\
 \texttt{\small lucio.marcenaro@unige.it} & \texttt{\small g.w.m.Rauterberg@tue.nl} & \texttt{\small carlo@dibe.unige.it} \\ \\ \\
 \end{tabular*}}

\address{
    \begin{tabular*}{0.7\textwidth}{c @{\extracolsep{\fill}}
c @{\extracolsep{\fill}}} $^1$ \small Department of Engineering (DITEN). &
\small  $^2$ Department of Industrial Design. \\ \small University of Genova &
\small Eindhoven University of \small Technology.\\ \small Genova, Italy &
\small Eindhoven, Netherlands.\\ \end{tabular*}}

\begin{abstract}
Wearable cameras allow people to record their daily activities from a user-centered (First Person Vision) perspective. Due to their favorable location, wearable cameras frequently capture the hands of the user, and may thus represent a promising user-machine interaction tool for different applications. Existent First Person Vision methods handle hand segmentation as a background-foreground problem, ignoring two important facts: i) hands are not a single "skin-like" moving element, but a pair of interacting cooperative entities, ii) close hand interactions may lead to hand-to-hand occlusions and, as a consequence, create a single hand-like segment. These facts complicate a proper understanding of hand movements and interactions. Our approach extends traditional background-foreground strategies, by including a hand-identification step (left-right) based on a Maxwell distribution of angle and position. Hand-to-hand occlusions are addressed by exploiting temporal superpixels. The experimental results show that, in addition to a reliable left/right hand-segmentation, our approach considerably improves the traditional background-foreground hand-segmentation.
\end{abstract}


\end{frontmatter}
\section{Introduction} \label{sec:intro}

The recent widespread availability of wearable devices has quickly attracted the interest of researchers, computer scientists and high-tech companies \cite{Starner2013}. The 90's idea of a body-worn device that is always ready to be used is nowadays possible, and its potential applicability to real problems is evident. In general, the wearable sensor that most attracted researchers' attention is the video camera: while enjoying a unique position to record what the user is seeing, it suffers from important issues and technical challenges \cite{Betancourt2014}. Images and videos recorded from this perspective are commonly referred to as First-Person Vision (FPV) or Egocentric videos \cite{Betancourt2014}.

\begin{figure}[h!]
	\centering
	\includegraphics[width=1\linewidth]{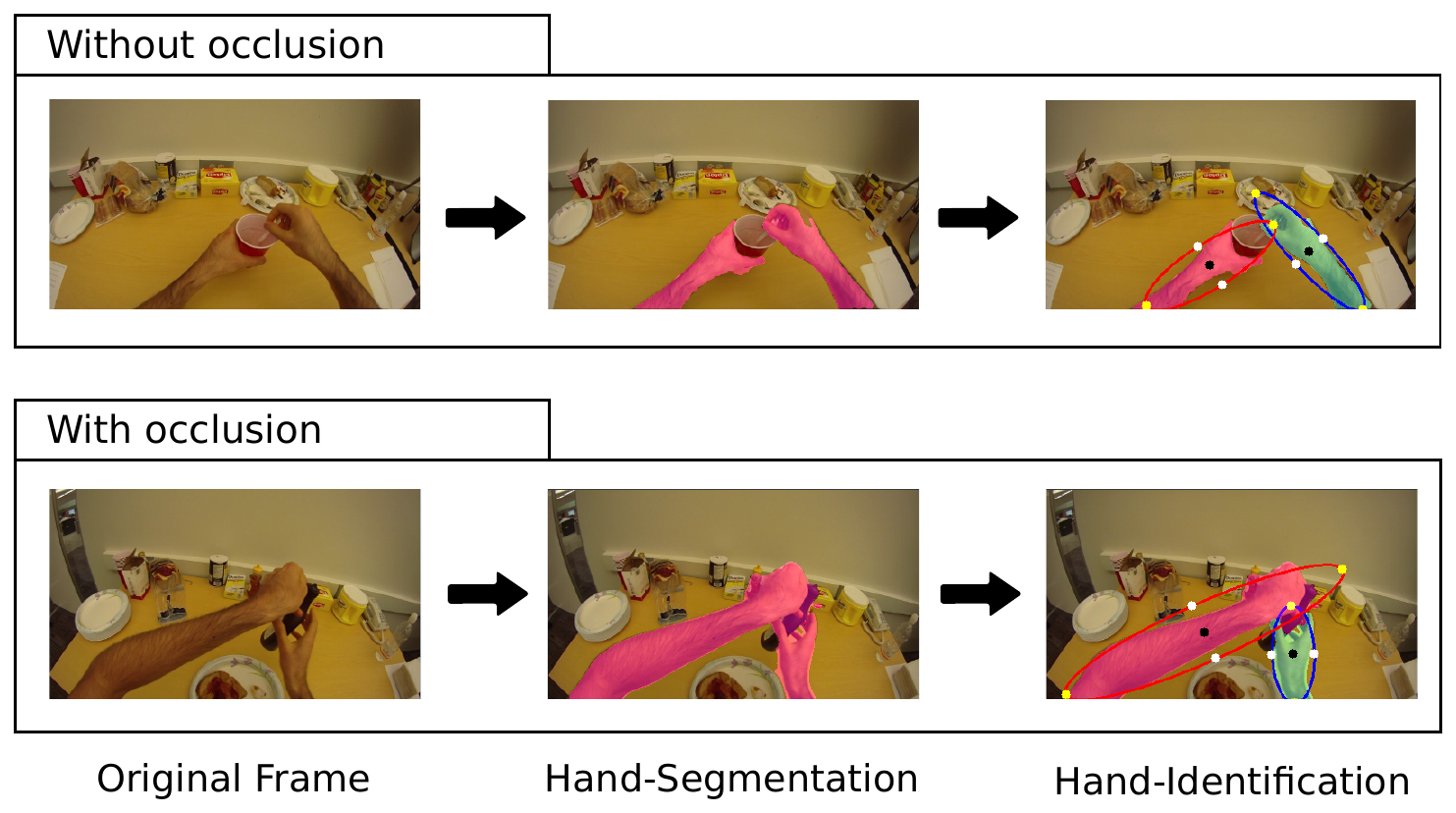}
	\caption{The difference between hand-segmentation and hand-identification}
    \label{fig:identification}
\end{figure} 

One of the more promising aspects of this video perspective is the tight link between the camera location and the user point of view, which makes it possible to frequently capture user's hands. A proper understanding of hand movements enables important applications such as activity recognition \cite{Fathi2011}, user-machine interaction \cite{Baraldi2015}, gaze estimation \cite{Fathi2012a, Buso2015}, hand-posture recognition \cite{Cai2015, Yang2015}, among others. The authors of \cite{Betancourt2015b} propose a hierarchical structure to develop hand-based methods and highlight several fields that might benefit of robust and efficient hand understanding techniques.

Hand-based studies are not restricted to wearable cameras and computer vision, in fact, biologists and neuroscientists have been deeply exploring hand usage in daily activities \cite{Renteria2012}, even before the emergence of modern wearable devices. There is a consistent number of studies and theories investigating hand-dominance in humans and its relationship with upper limb motion skills. It is estimated that $9$ out of $10$ individuals are right-handed and as a consequence their upper limb skills are asymmetric for what concerns speed, control and strength. Interestingly, these findings are similar across different geographic locations and cultures \cite{Mcmanus2009}. 

Current FPV hand-based literature consistently approaches inference over hands as a background/foreground segmentation problem, where hand-like pixels represent the foreground while the remaining pixels define the background \cite{Li2013b, Fathi2011a}. Even if this approach provides a broad range of algorithmic approaches and evaluation criteria based on machine learning and computer vision, it oversimplifies the biological perspective, ignoring hand-dominance and limiting the capabilities of wearable cameras to understand hand interactions and asymmetric upper limb motor skills. Figure \ref{fig:identification} shows the difference between the traditional binary hand-segmentation and a left/right(L/R) hand-segmenter as proposed in this work. The second row shows an example on which both hands interact closely thus causing a hand-to-hand occlusion. The standard binary hand-segmenter fails in explaining these occlusions and creates a single hand-like segment while the L/R hand-segmenter can detect and split two hands correctly.

\begin{figure}
    \centering
    \subfigure[Asymmetry in the position of the hands.]{
    \includegraphics[width=0.27\linewidth]{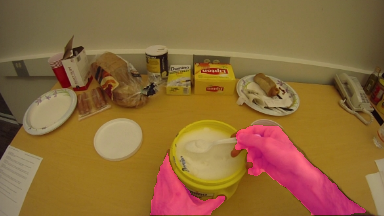}\label{fig:asymmetry}
    }\hfill
    \subfigure[Manipulating objects in the borders of the frame.]{
    \includegraphics[width=0.27\linewidth]{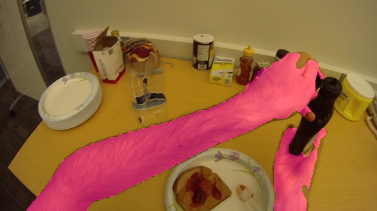}\label{fig:borders}
    }\hfill
    \subfigure[Hand-Segmentation with Occlusion problem.]{
    \includegraphics[width=0.27\linewidth]{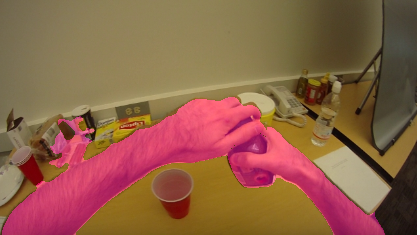}\label{fig:occlusion}
    }
    \caption{Hand-Segmentation examples.}
    \label{fig:segmentation}
\end{figure}

Concerning hand-identification, some authors have suggested a strong relation between hand identity and the position of hand-like segments \cite{Philipose2009}. Intuitively, segments located on the left, or right side of the frame belong to the left or right hand respectively. Figure \ref{fig:segmentation} shows some examples of frames where this approach does not work. In summary, the location-based strategy performs well if the symmetry between the hands is almost perfect and there are not hand-to-hand occlusions.

From our experience, the capability of wearable cameras to distinguish one hand from the other is critical. This is particularly true when it comes to understand bi-manual tasks \cite{Swinnen2004, Vincze2009}, e.g. in medical rehabilitation of upper-limb stroke \cite{Turolla2013} and cerebral palsy \cite{Speth2013a}. Another field that can benefit from this independent understanding is neuroscience, where hand-dominance is commonly associated with several neurological factors \cite{Goble2008}. For example, the authors of \cite{Knaus2016, Cook2013} found significant differences in the hand-dominance level of children with Autism Spectrum Disorder (ASD). A wearable camera that is able to differentiate between left and right hand is not only in line with biological and neurosciences perspective, but it also opens the doors to understanding hands as two interacting entities, centrally coordinated to achieve a particular goal.

To the best of our knowledge, this is the first work exploring in detail a L/R hand-segmentation, considering realistic scenarios with hand-to-hand occlusions and asymmetric hand positions. The contribution of this paper is three-folded: i) It proposes a theoretical hand-identification model based on Maxwell distributions to decide whether a hand-like segment corresponds to a left or a right hand ii) It faces the hand-to-hand occlusions problem by exploiting temporal superpixels under a dynamic procedure. iii) It significantly improves a state-of-the-art binary hand-segmenter by using a multi-model classification strategy and assuming that one left and one right hand appear in front of the user at most. The last assumption is valid, especially in the aforementioned medical applications, where no human-to-human interaction is required \cite{Turolla2013}. 

The remaining of this paper is organized as follows: Section \ref{sec:stateoftheart} summarizes the state of the art in hand-based methods in FPV. In section \ref{sec:method} our approach is presented and subsequently evaluated in \ref{sec:results}. The provided evaluation is performed sequentially by first analyzing each components independently. Finally, \ref{sec:conclusions} concludes and provides future research lines based on the obtained result.

\section{State of the Art}\label{sec:stateoftheart}

Recent literature \cite{Betancourt2014} highlights the significant role of hands in FPV. Several promising hand-based applications are frequently mentioned, such as activity recognition \cite{Fathi2011} and user-machine interaction \cite{Baraldi2015}, among others. Different authors have also sketched other advanced and more realistic applications. However, real applicability is still restricted by the limited capabilities of current methods to work under realistic conditions, such as illumination changes, or complex hand interactions\cite{Betancourt2015b}.


In \cite{Betancourt2015b} the authors propose a unified structure for hand-based methods, which highlights the importance of understanding hands at different hierarchical levels (i.e. hand-detection, hand-segmentation, hand-identification and hand-tracking). The first level of the structure is hand-detection, which answers the yes-no question about hand presence in a frame. The objective of this level is to optimize computational resources, and to reduce the false positives rate when hands are not being recorded by the camera \cite{Betancourt2014a,Betancourt2015a}. Hand-detection is commonly faced as a frame-by-frame classification problem \cite{Betancourt2015}, and is frequently given as granted when studying controlled tasks where the user is always manipulating objects in front of the camera, for example in the Kitchen \cite{Fathi2012a} and the EDSH \cite{Li2013a} datasets.

Once hands are detected, the following step, and most studied level, is hand-segmentation. The goal of hand-segmenters is to find the set of pixels of a particular frame belonging to the hands of the user. Recent hand-segmenters can be considered advanced implementations of the color-based seminal work of \cite{Jones1999}. Remarkable results are obtained in the work of \cite{Li2013a} where a Random Forest classifier is trained to discriminate positive and negative pixels. That work also explores the use of texture and the fusion of multiple hand-segmenters to deal with changing light conditions \cite{Li2013b}. This strategy is further improved in \cite{Zhu2014} by preserving the shape of the hands using a shape-aware classifier and also in \cite{Baraldi2015} by using superpixels. The authors in \cite{Zhu2015} use the segmented hands to divide the hand-like segments in fingers, palm, and arm.

According to the framework proposed in \cite{Betancourt2015b}, hand-identification is more than an incremental step towards the solution of the problem, since it opens a range of new possibilities and applications. It makes technically possible to make a paradigm shift towards viewing the hands as two interacting entities working jointly to achieve a particular goal. Literature on hand-identification is insufficient: the problem is usually regarded as a post-processing step performed after segmentation. Ren et al. \cite{Philipose2009}, for instance, use the side of the frame where the skin-like segments are located to label it as the left or right hand. We can identify three common cases in which this approach does not work correctly: i) The symmetry of the skin-like segments is affected by changes in the attention point or the camera position, Figure \ref{fig:asymmetry}. ii) The user is manipulating an object close to the frame borders, Figure \ref{fig:borders}. iii) The hands are close enough to be segmented as a single skin-like segment (hand-to-hand occlusion), Figure \ref{fig:occlusion}. 

To address these cases, Fathi et al. \cite{Fathi2011a} use a Support Vector Machine that is able to classify each frame into four categories, i.e. single left hand, single right hand, two different hands, two interacting hands. These categories oversimplify the hand-identification problem since they do not provide a L/R hand-segmentation. In the same line of research, \cite{Buso2015} extends the approach of \cite{Fathi2011a} by using the relative positions of the segmented hands and the active objects to build a goal oriented model of attention. Denoted also as a hand-identification but targeting a different purpose, the authors of \cite{Lee2014} propose a Bayesian method to identify if a hand-like segment belongs to the user or to somebody in front of him. This problem clearly provides an alternative definition of hand-identification, which is particularly important when the user is interacting with other people. The authors illustrate the importance of this approach by using a dataset recorded for medical experiments with children.

The primary goal of the present work is to address the hand-identification problem following the definition proposed in the independent studies \cite{Betancourt2014} and \cite{Philipose2009}. Our approach relies on a multi-model implementation of the binary hand-segmenter proposed by \cite{Li2013a, Li2013b}, but can be easily extended to future improvements of any hand-segmentation algorithm. Compared to the state of the art literature, we highlight three important novelties:

\begin{itemize}
    \item{The proposed L/R hand-segmenter significantly improves segmentation score of the state-of-the-art by fusing multiple Random Forests to capture light changes and by exploiting the fact that the user has at most one left and one right hand. This assumption is particularly useful when studying bi-manual tasks in controlled environments. The experimental results show that, in addition to the reliable left and right information, our final segmentation improves the state-of-the-art binary hand-segmenter \cite{Zhu2014} of around $10$ $F1$ score points in some videos of the kitchen dataset \cite{Fathi2011a}.}
    \item{In contrast to \cite{Fathi2011a}, our approach relies on simple set algebra to detect occlusions, which is computationally more efficient and achieves a detection level of $99\%$ (as shown in section \ref{sec:results}). Moreover, our method not only provides a category label but splits the occluded binary hand-segment by using superpixels.}
    \item{Given a previous occlusion detection and split, we propose a probabilistic L/R hand-identification model using a max-likelihood ratio test of two Maxwell distributions based on position and angle. Our approach is robust to asymmetries in the hand positions and can be tuned for different camera locations and lenses. The experimental results show that our method accurately identifies $99\%$ of the manual masks of the kitchen dataset \cite{Fathi2011a}.}
\end{itemize}

\section{Our approach}\label{sec:method}

\begin{figure}[h!]
	\centering
	\includegraphics[width=1\linewidth]{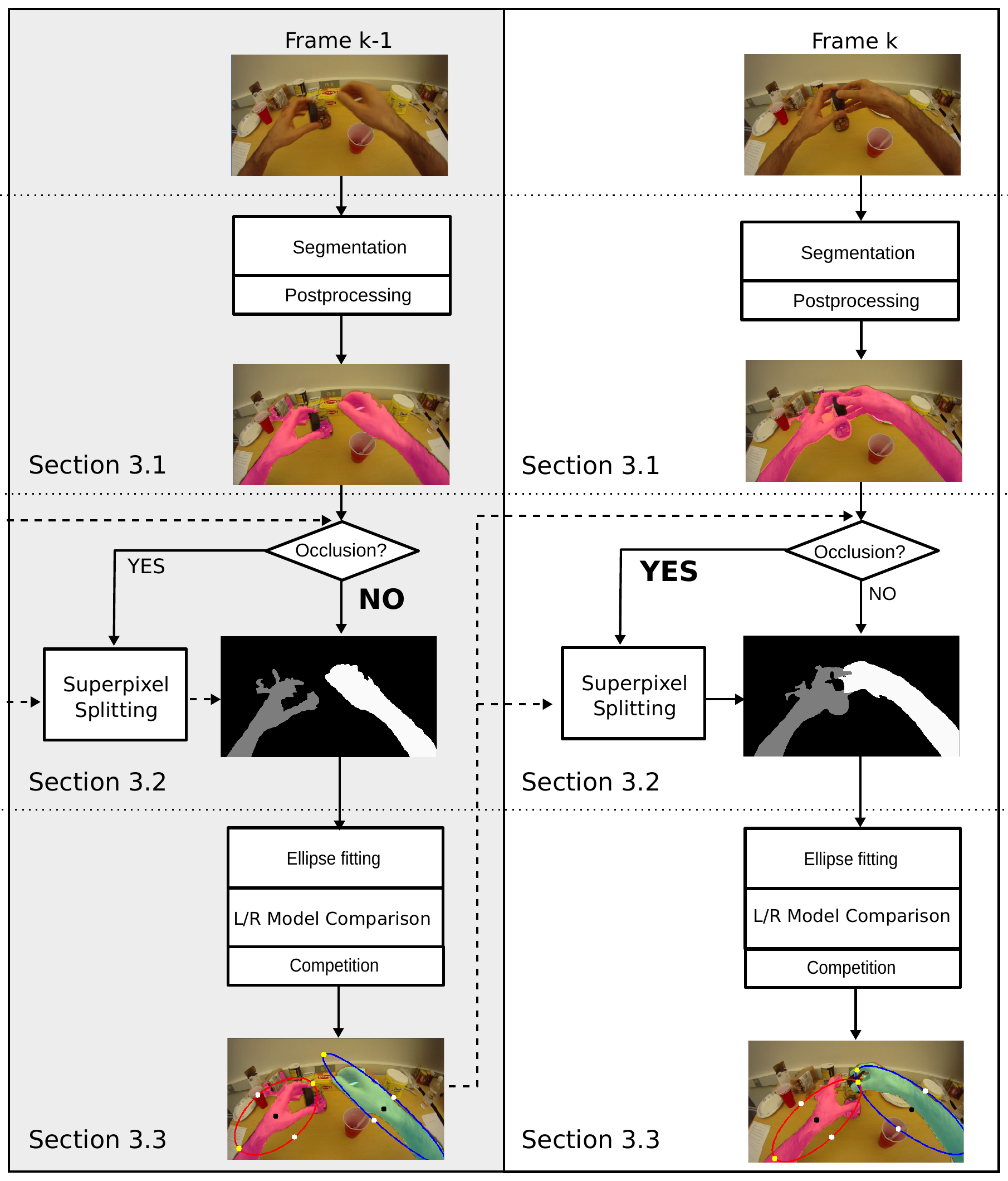}
	\caption{General description of our approach}
		\label{fig:block_diagr}
\end{figure}

Our final goal is to extract an accurate L/R hand-segmentation that is robust to hand-to-hand occlusions, asymmetric hand configurations and object manipulations in the borders of the frame. Figure \ref{fig:block_diagr} summarizes the proposed work-flow; at the top of the diagram there is the input frame, while the resulting L/R hand-segmentation is shown at the bottom. The intermediate stpdf are the hand-segmentation (section \ref{sec:handSegmentation}), the hand-to-hand occlusion detection and disambiguation (section \ref{sec:occlusionStep}), and the hand-identification (section \ref{sec:LRmodel}). These stpdf are in line with the unified structure proposed in \cite{Betancourt2015b}. It is important to note that the three levels (i.e. Hand-Segmentation, Occlusion-Detection, Hand-Identification) are mutually independent; which makes it possible to improve them separately or using more complex sensors instead. As an example, the occlusion detector can be applied on hand-segments coming from a RGBD cameras, and the hand-identification method can be used on top of a faster occlusion detector, or even directly on the hand-segmenter if the occlusions are not relevant for a particular application.


From the diagram it can be noticed that, at each time instant the procedure exploits the previous L/R segmentation to detect and split the hand-to-hand occlusions. This temporal dependency requires a reliable previous L/R segmentation and no hand-to-hand occlusion in the initial frame. Intuitively, the higher the sampling rate, the more reliable the occlusion detection; however, as will be shown in \ref{sec:results}, even using sampling rates of $15FPS$ the final segmentation accuracy of the left and right hand is still around $90\%$. The main goals of this work are the algorithmic performance and the segmentation capabilities. To reach real-time performance, the algorithm must be optimized by balancing the compression width, the sampling rate, or by developing parallel versions of the Random Forest and/or the superpixel algorithm. Our current work points out that, using GPU implementation and an image resampler (outputting $600$ pixels width images and preserving aspect-ratio), it is possible to achieve a throughput of $30FPS$.


\subsection{Binary Hand-Segmentation}\label{sec:handSegmentation}

\begin{figure}[h!]
	\centering
	\includegraphics[width=1\linewidth]{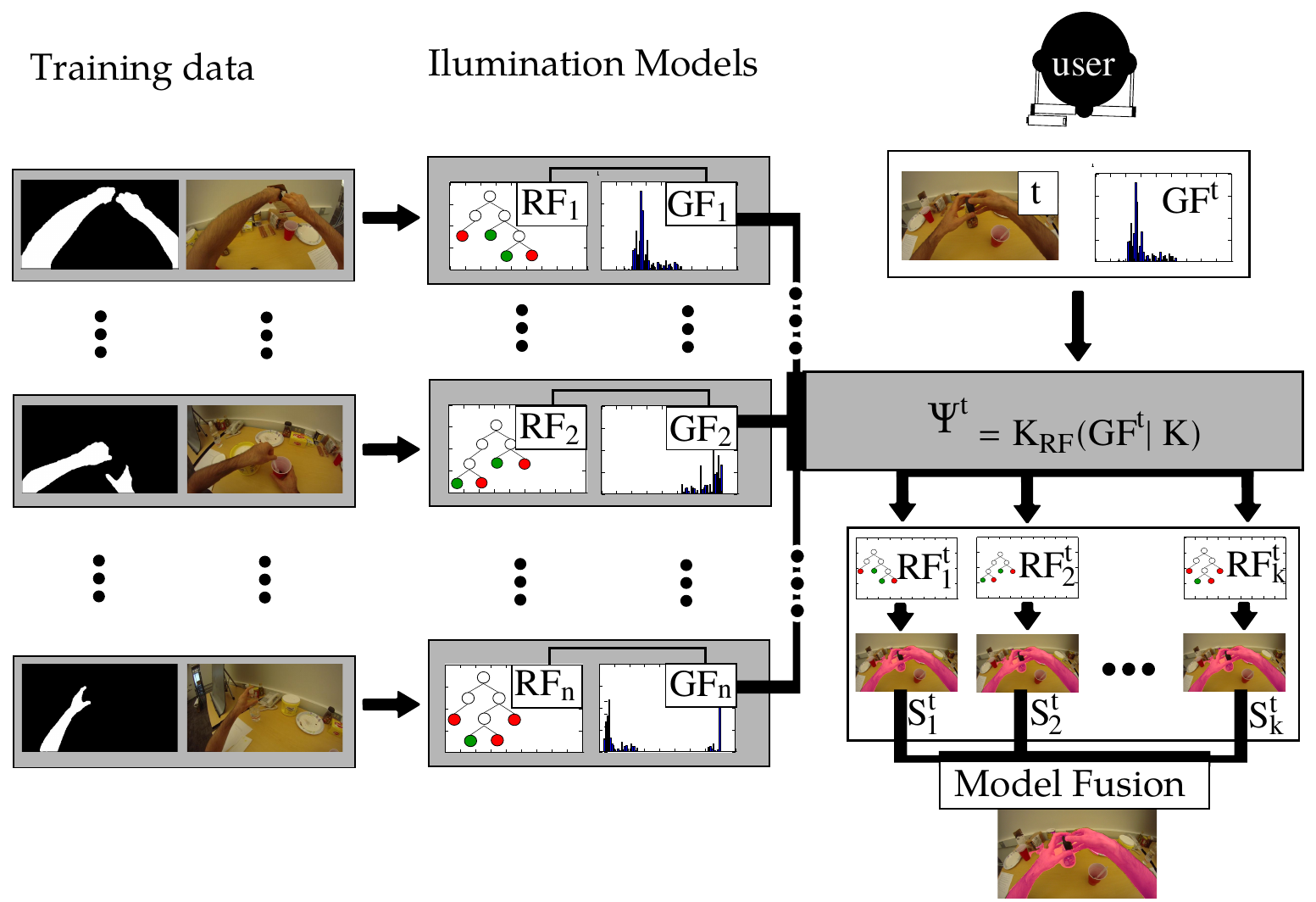}
	\caption{Multi-model binary hand-segmenter}
		\label{fig:multimodel}
\end{figure}

At this level there is no difference between left and right hand. The objective is to discriminate pixels of the frame that looks like the hand-skin based on color. This level is based on a multi-model version of the pixel-by-pixel binary hand-segmenter proposed by \cite{Li2013b}. Figure \ref{fig:multimodel} summarizes the general idea of the multi-model approach. The gray blocks correspond to the training while the white blocks to the testing.

The first column of the figure contains the manual masks and their corresponding raw frames. The masks were extracted using the graph cut manual segmenter provided by \cite{Li2013b}. Let us denote $N$ as the number of manual masks available in the dataset, and $n$ as the number of training pairs selected to build a multi-model binary hand-segmenter. For each training pair $i=1\dots n$ a trained binary random forest ($RF_i$) and its global features ($GF_i$) are obtained and stored in order to construct a pool of illumination models (second column of the figure).  Each $RF_i$ is trained using as features the $LAB$ values of each pixel in the frame $i$ and as class their corresponding values in the binary masks. As global feature ($GF_i$) we use the flatten HSV histogram. The choice of the color spaces is based on the results reported by \cite{Li2013b} and \cite{Morerio2013}. Once the illumination models are trained, a K-Nearest-Neighbors structure, denoted as $K_{RF}$, is estimated using as input the global features $GF_i$.

In the testing phase, the $K_{RF}$ is used as a recommender system which maps the global features frame to the indexes of the nearest $K$ illumination models ($RF^t$). These are used to obtain $K$ possible segmentations ($S^t$), which are eventually fused to obtain the final hand-segmentation ($HS^t$). This procedure is illustrated in the third column of Figure \ref{fig:multimodel}. Formally, let's denote the testing frame as $t$ and its HSV-histogram as $GF^{t}$, the indexes of the closest $K$ illumination models ordered with increasing euclidean distances as equation (\ref{eq:indexes}), their corresponding $K$ random forest as equation (\ref{eq:models}), and their pixel-by-pixel segmentation applied to $t$ as equation (\ref{eq:segmenters}).

\small
\begin{eqnarray}
    \Psi^{t} & = & K_{RF}(GF^t|K) \label{eq:indexes} \\ 
             & = & \{\psi_1^t,\dots,\psi_K^t\} \nonumber \\ \nonumber \\
    RF^t & = & \{RF_{\psi_1^t},\dots,RF_{\psi_K^t}\} \label{eq:models} \\ \nonumber \\
    S^t & = & \{RF_{\psi_1^t}(t),\dots,RF_{\psi_K^t}(t)\} \label{eq:segmenters} \\
             & = &  \{S_1^t,\dots,S_K^t\} \nonumber
\end{eqnarray}
\normalsize


The binary hand-segmentation of the frame is the normalized weighted average of the individual segmentations in $S^t$, which is formally given by equation (\ref{eq:fussion}); $\lambda$ is a decaying weight factor that is equal to $0.9$, based on the results of \cite{Li2013b}. The weights $S^t$ are then set as $\{0.9,0.9^2,0.9^3 \cdots 0.9^K\} = \{0.9,0.81,0.729 \cdots 0.9^K\}$. With this in mind, the hand-segmentation has 2 parameters to be defined, namely the number of illumination models ($n$) and the number of closest random forests to average ($K$). These parameters are defined in section \ref{sec:results} following an exhaustive evaluation in the Kitchen dataset. 

\small
\begin{eqnarray}
    HS^t & = & \frac{\sum_{j=1}^{K}\lambda^{j}\cdot S^t_j}{\sum_{j=1}^{K}\lambda^{j}} \label{eq:fussion}
\end{eqnarray}
\normalsize

At this point, there is a set of hand-like segments, some of them matching the hands of the user (true-positives) and some of them as the result of pixels in the image with similar color to the user skin (false-positives). If a fixed camera location (e.g. head, chest, shoulder) is known, then it is possible to define a set of post-processing rules to remove some of the false-positives. The post-processing has $4$ stpdf: i) Find the contours (polygons) containing the hand-like pixels; ii) Remove those contours far of the left, bottom or right margin; iii) Remove the contours smaller than $0.1 \cdot w^2$, where $w$ is the width of the frame; iv) keep the largest $3$ contours. We perform an extra filtering stage after the hand-identification step to keep only the best left and best right contour. 

\subsection{Hand-to-hand Occlusions}\label{sec:occlusionStep}

\begin{figure}[!ht]
	\centering
	\includegraphics[width=1\linewidth]{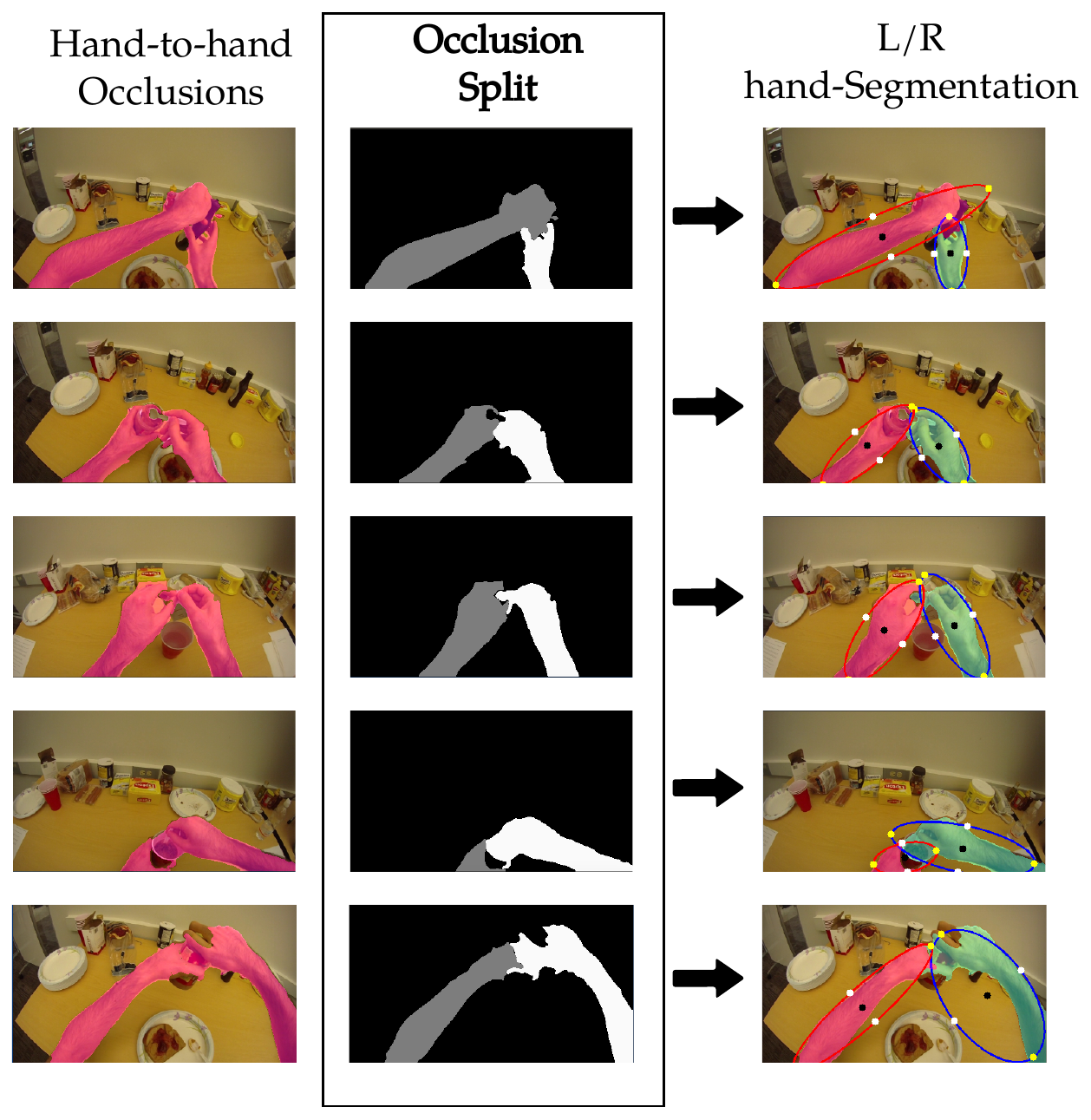}
	\caption{Examples of Hand-to-hand occlusion split.}
	\label{fig:nosplit}
\end{figure}

The proposed L/R hand-identification model assumes that the hand-like segments are not occluded or have been split before. If hand-to-hand occlusions were ignored the L/R hand-identification model would process a larger hand-like segment and would assign it completely as left or right. Moreover, ignoring the occlusions would make the tracking of the hands more complex due to frequent flickering in the hand-identification. To avoid these cases we perform an occlusion detection step (section \ref{sec:occlusionDetection}) followed by a segmentation split (section \ref{sec:occlusionDisambiguation}). Figure \ref{fig:nosplit} shows some examples of occlusion (first column) and the split (second column). The third column shows the result of the L/R hand-segmentation if occlusions are properly handled.

\subsubsection{Hand-to-hand Occlusion Detection}\label{sec:occlusionDetection}

The main goal of this step is to decide whether the hand-like segments of a particular frame come from a hand-to-hand occlusion. Given a reasonable sampling rate, it is possible to assume that a hand-to-hand occlusion requires the presence of both hands in the previous. Let's define then, the previously detected L/R hand-segments, as $L^{t-1}$ and $R^{t-1}$ respectively, and the larger binary hand-segment of the current frame as $HS_i^t$. Two important assumptions must be verified here: i) There is not hand-to-hand occlusions in the first frame of the video segments containing both hands, ii) The detection and split reliability is high. The first assumption is particularly true for videos recording realistic hand interactions like the kitchen dataset. The second assumption is evaluated in section \ref{sec:perfectSegmentationResults} and \ref{sec:occlusionDissambiguationResults}.

In case of occlusion, and given a small sampling rate, $HS_i^t$ must intersect simultaneously with $L^{t-1}$ and $R^{t-1}$. If this happens, it can be assumed that the hands are close enough, or connected by noisy pixels, to be considered a hand-to-hand occlusion. Algorithm \ref{alg:occlusion} formally defines the hand-to-hand occlusion detection. For the sake of the compactness of notation, we use a bar over segments to refer to their area (e.g. $\bar{S}^{t-1}$ refers to the area of segment $S^{t-1}$). As it can be seen in the pseudocode, $HS_i^t$ is defined as hand-to-hand occlusion if the area of its intersection with $L^{t-1}$ and $R^{t-1}$ is between $80\%$ and $120\%$ of the total area of the L/R hand-segments. 

\begin{algorithm}
\caption{Hand-to-hand occlusion detection.}
\begin{algorithmic}[1]
\scriptsize
\Procedure{IsOcclusion}{$HS_i^t, L^{t-1}, R^{t-1}$}
\If{$L^{t-1} \neq \varnothing \textsc{\textbf{ AND }} R^{t-1} \neq \varnothing$}
    \State $S^{t-1} \gets L^{t-1} \cup R^{t-1}$ 
    \State $I \gets HS_i^t \cap S^{t-1}$  \vspace{2pt}
    \If{$0.8 \cdot \bar{S}^{t-1} \le \bar{I} \le 1.2 \cdot \bar{S}^{t-1}$}
        \State $Occlusion  \gets  True$
    \Else
        \State $Occlusion  \gets  False$
    \EndIf
\Else
    \State $Occlusion  \gets  False$
\EndIf \\
\Return $Occlusion$
\EndProcedure
\end{algorithmic}
\label{alg:occlusion}
\end{algorithm}

\subsubsection{Occlusion splitting}\label{sec:occlusionDisambiguation}

In the case of hand-to-hand occlusion, the next step is to split the affected segment in two parts by exploiting its inner edges and the previous L/R hand-segments. Following the notation of section \ref{sec:occlusionDetection}, let us define $\Phi^t = \{\phi_0^t, \phi_1^t, \dots, \phi_k^t \}$ as a superpixel representation of the frame $t$. Pseudocode \ref{alg:split} summarizes the stpdf to hand-to-hand occlusions split. Our approach initially relies on the intersection with the previous L/R segments and subsequently, if necessary, in the superpixels of the previous frame. 

\begin{algorithm}
\caption{Splitting occluded hand segments.}
\label{alg:split}
\begin{algorithmic}[1]
\scriptsize
\Procedure{SplitOcclusion}{$BigHandBlob, L^{t-1}, R^{t-1}, \Phi^{t-1}, \Phi^{t}$}
\State $segment_1 \gets \varnothing$
\State $segment_2 \gets \varnothing$
\For{ $\phi_i^t \in \{ \Phi^{t} \mid \phi_i^t \in BigHandBlob \}$ } \vspace{2pt}
    \If{$centroid(\phi^t_i) \in L^{t-1}$}
        \State $segment_1 \gets segment_1 \cup \phi_i^t$
    \ElsIf{$centroid(\sigma^t_i) \in  R^{t-1}$}
        \State $segment_2 \gets segment_2 \cup \phi_i^t$
    \Else
        \State $tmp \gets \{ \phi \in \Phi^{t-1} \mid centroid(\phi_i) \in \{L^{t-1} \cup R^{t-1} \}\}$
        \State $closest_\phi  \gets \argminl_{\phi \in tmp}\| \phi - \phi_{i}^{t} \|^2$ \label{alg:clstSuperpixel} \vspace{2pt}
        \If{$centroid(closest_\phi) \in L^{t-1} $}
            \State $segment_1 \gets segment_1 \cup \phi_i^t$
        \Else
            \State $segment_2 \gets segment_2 \cup \phi_i^t$
        \EndIf
    \EndIf
\EndFor\\
\Return{$segment1, segment2$}
\EndProcedure
\end{algorithmic}
\end{algorithm}

Intuitively, the intersection of current hand-segments with the previous L/R segments provides reliable decisions for small sampling rates. However, due to the fast camera and hand moves, not all the pixels inside the occluded hand-segment can be solved in this way. For these pixels, we rely on the closest previous superpixel. The higher the sampling rate the more relevant the superpixel criteria.

In practice, we use the original SLIC algorithm as the superpixel method with the metric defined by (\ref{eq:metric}). The same metric is used to find the closest previous superpixel in algorithm \ref{alg:split} line \ref{alg:clstSuperpixel}, where $\| \phi_i - \phi_j \|^2_c$ is the color metric given by equation (\ref{eq:colorMetric}), $\| \phi_i - \phi_j \|^2_s$ is the space metric given by equation (\ref{eq:spatialMetric}), and $m^2$ is the spatial weight.  Our experimental results use $LAB$ as color space for two reasons: i) It has been pointed out as the best performing feature for hand-segmentation in egocentric videos ii) It is the feature used in the original SLIC algorithm. \footnote{See \cite{Alata2009} for a detailed comparison of different color spaces and their discriminative power.}

\small
\begin{eqnarray}
    \| \phi_i - \phi_j \|^2 &=& \| \phi_i - \phi_j \|^2_c + m^2 \| \phi_i - \phi_j \|^2_s \label{eq:metric} \\
    \| \phi_i - \phi_j \|_c &=& \sqrt{(l_i - l_j)^2 + (a_i - a_j)^2 + (b_i - b_j)^2} \label{eq:colorMetric} \\
    \| \phi_i - \phi_j \|_s &=& \sqrt{(x_i - x_j)^2 + (y_i - y_j)^2} \label{eq:spatialMetric}
\end{eqnarray}
\normalsize

\subsection{Hand Identification}\label{sec:LRmodel}

Assuming that the current frame is not occluded, or has been previously split, the next step is to decide if detections are left or right hands. As explained before, using only the horizontal position of the hands in the frame is not always reliable. It can be hypothesized that, by extending the horizontal position with the hand orientation, it is possible to the solve the difficult situations. To confirm this, it was performed an exhaustive analysis of the kitchen hand-masks extended with labels about the hand identity. These masks are subsequently used to define a probabilistic L/R hand-identification model based on the best-fitting ellipses (section \ref{sec:model}). The ellipses are fitted with the algorithm proposed by \cite{Fitzgibbon1995}.  Finally, the already mentioned Maxwell model is used in a likelihood ratio test to exploit the fact that one left/right hand can be present at most (section \ref{sec:likehoodTest}). It is noteworthy that the proposed identifier does not need initialization: it is independent of the sampling rate, and can be applied to frames with left, right, or both hands. Additionally, its parametric nature opens the door for further integration for higher inference levels as proposed in \cite{Betancourt2015b}.


\subsubsection{Building the L/R hand-identification model}\label{sec:model}

\begin{figure} [h!]
    \centering
    \subfigure[{\scriptsize Geometric problem of the left hand-segment}]{
    \includegraphics[width=3.8cm]{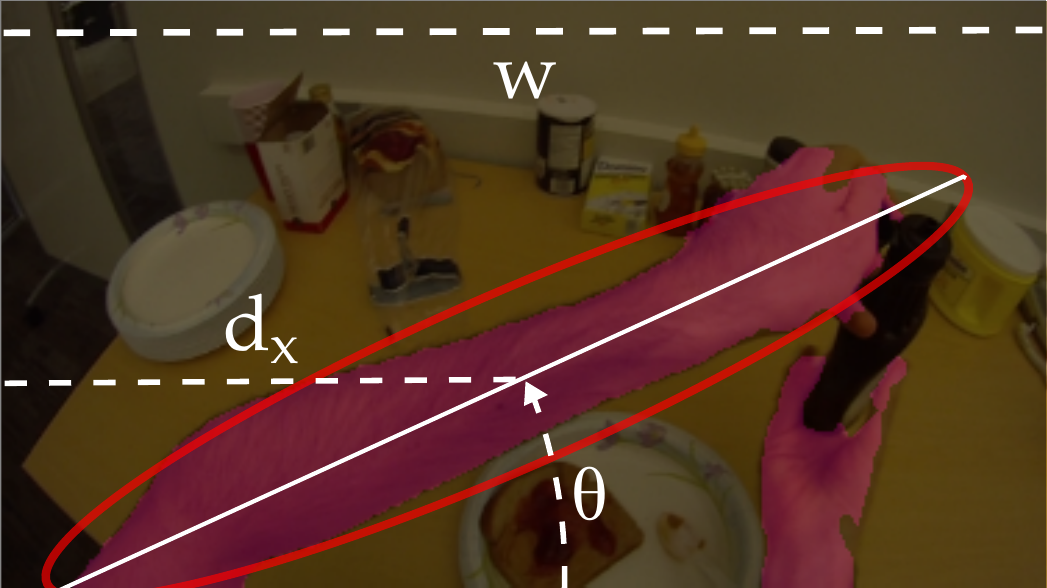}\label{fig:left_ellipse}
    }
    \subfigure[{\scriptsize Geometric problem of the right hand-segment}]{
    \includegraphics[width=3.8cm]{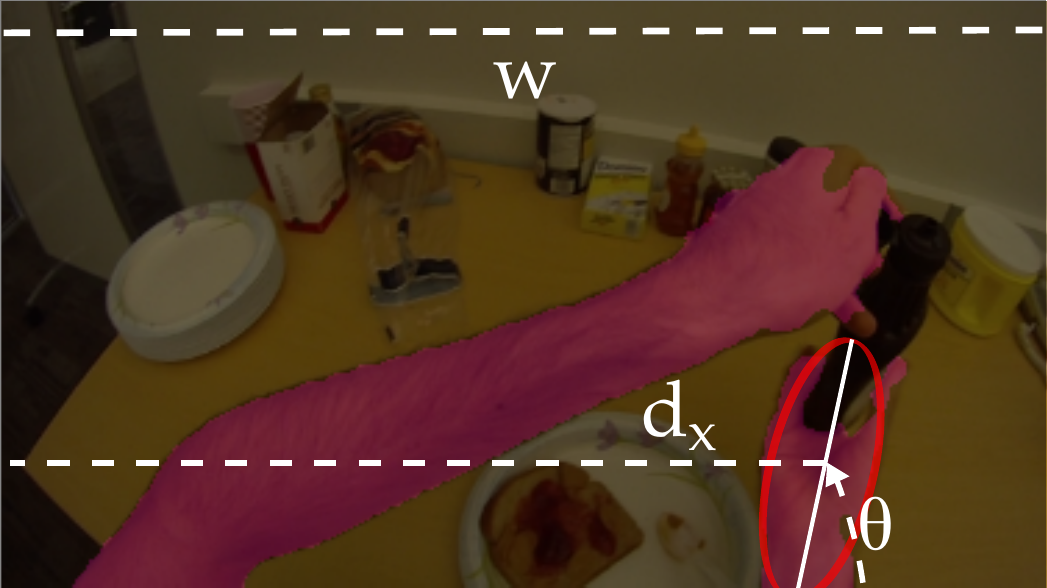}\label{fig:right_ellipse}
    }
    \caption{Input variables for the L/R hand-identification model.}
    \label{fig:geometricProblem}
\end{figure}

A quick analysis of egocentric videos of daily activities easily points to the angle of the hands with respect to the lower frame border ($\theta$), and the normalized horizontal distance to the left border ($x$) as two discriminative variables to build our L/R hand-identification model. Figure \ref{fig:geometricProblem} illustrates these variables. For the remaining part of this section $x$ is the normalized value of $d_x$ with respect to the frame width $w$.

\begin{figure}[!ht]
	\centering
	\includegraphics[width=1\linewidth]{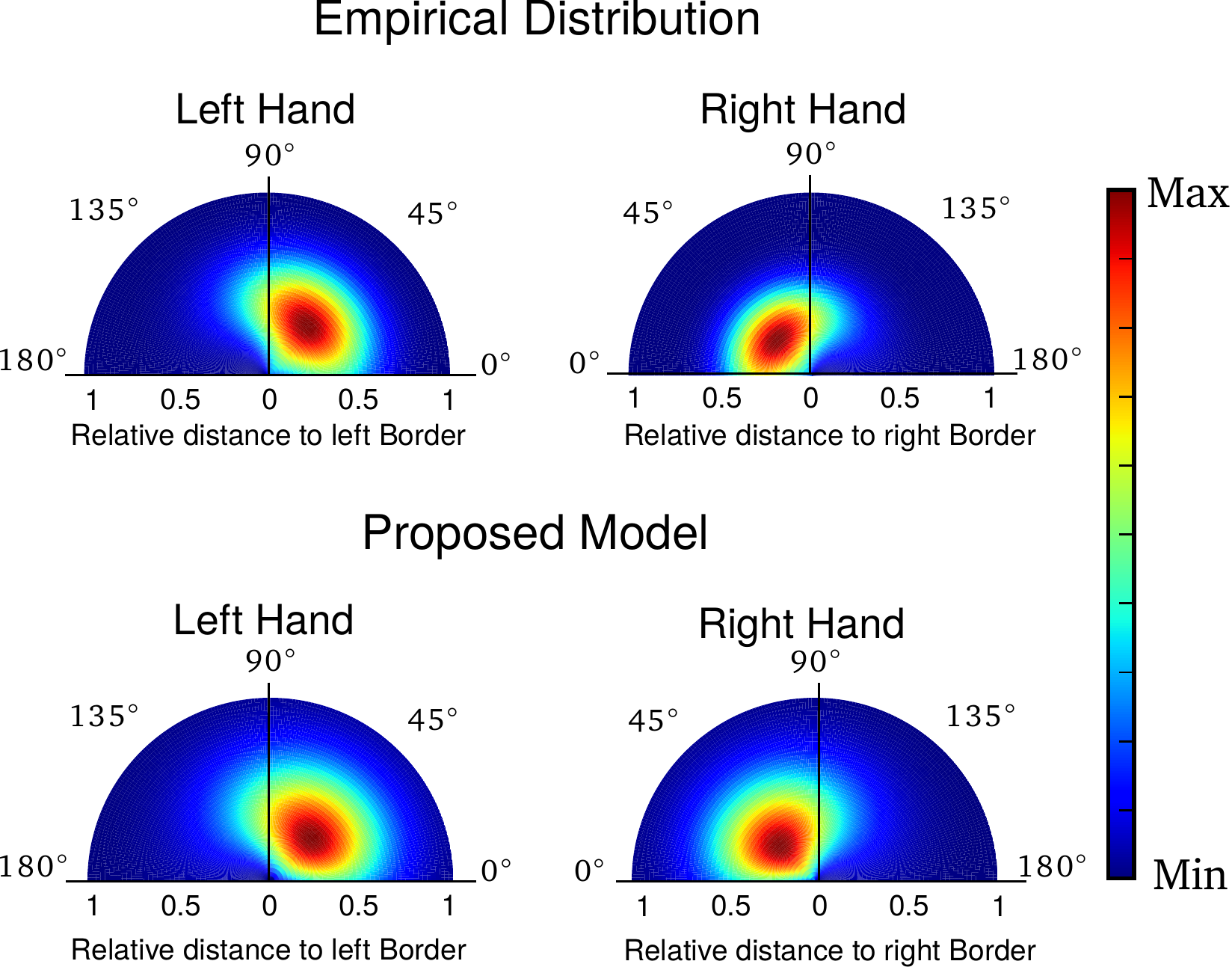}
	\caption{Empirical (\textbf{Top}) and theoretical (\textbf{Bottom}) hand distribution function given the distance to relative distance to the sides of the image. For the left(right) the relative distance to the left(right) side is used.} \label{fig:idProbabilities}
\end{figure}

The upper half of Figure \ref{fig:idProbabilities} shows the observed empirical distribution of the $x$ and $\theta$ for the left and the right hands of the kitchen dataset extended masks.  In the horizontal axis is the relative distance to the left border ($x$), for the left hand-like segments, and the relative distance to the right border ($1-x$), for the right hand-like segments. The angular dimension is the anti-clockwise angle with respect to the horizontal border of the frame ($\theta$). Interestingly, there is a small asymmetry between the left and right distributions, meaning that one of the two hands is used for a wider variety of movements than the other. We point this as an interesting finding that could lead to further device personalization depending on the dominant hand of the user, or to analyze the hand usage in daily activities.

Based on the empirical distributions, a mathematical formulation to fit the observed distribution is proposed; which, interestingly can be easily approximated two independent Maxwell distributions. The reasons behind the choice of the Maxwell distribution are the following: i) It is positive defined ii) It allows to include an asymmetry factor in our formulation. The mathematical formulation for the left hand ($p_l$) and the right hand ($p_r$) is given by equation (\ref{eq:distr1}) and (\ref{eq:distr2}) respectively, where $p_x$ is the Maxwell distribution with parameters $\Theta = [d,a]$. The values of $x$ and $\theta$ are defined in the interval $[0,1]$ and $[0,\pi]$. In general $d$ controls the displacement of the distribution (with respect to the origin) and $a$ controls its amplitude. 

\small
\begin{eqnarray}
	\label{eq:distr1}
	p_{l}(x, \theta|\Theta_l^x,\Theta_l^{\theta}) &=& p(x|\Theta_l^x)  p(\theta|\Theta_l^{\theta}) \\
	\label{eq:distr2}
	p_{r} (x,\theta|\Theta_r^x,\Theta_r^{\theta}) &=&  p(1-x|\Theta_r^x)  p(\pi-\theta|\Theta_r^{\theta}) \\ \vspace{30pt}
	p(x|\Theta) = p(x|d,a) &=& \sqrt{\frac{2}{\pi}}\dfrac{(x-d)^2}{a^3} \: e^{-\dfrac{(x-d)^2}{2a^{2}}} \label{eq:max_horizontal}
\end{eqnarray}
\normalsize

In total, our formulation contains $8$ parameters summarized in equation (\ref{eq:parameters}). As notation, the subscript of $\Theta$ refers to the left ($l$) or right ($r$) parameters, and the superscript refers to the horizontal distance ($x$) or the anti-clockwise angle ($\theta$). The parameters of the model are selected by fitting the empirical distribution and the final values are given by equation (\ref{eq:tunning}). The second row of Figure \ref{fig:idProbabilities} shows the theoretical distribution.

\small
\begin{eqnarray}     \begin{bmatrix}
        \Theta_l^x & \Theta_l^{\theta} \\ 
        \Theta_r^x & \Theta_r^{\theta} \\ 
    \end{bmatrix}
        & = &
    \begin{bmatrix}
        d_l^x & a_l^x & d_l^{\theta} & a_l^{\theta}  \\ 
        d_r^x & a_r^x & d_r^{\theta} & a_r^{\theta}  \\ 
    \end{bmatrix} \label{eq:parameters}\\
        & = &
    \begin{bmatrix}
        -0.05 & 0.24 & -0.63 & 0.94  \\ 
        -0.08 & 0.21 & -0.91 & 1.10  \\ 
    \end{bmatrix} \label{eq:tunning}
\end{eqnarray}
\normalsize

\subsubsection{Using the L/R hand-identification models}\label{sec:likehoodTest}

To compare the fitting performances of the L/R hand-identification models given by equation (\ref{eq:distr1}) and (\ref{eq:distr2}), a likelihood ratio test on the post-processed hand-like segments is performed. The likelihood ratio test is given by equation (\ref{eq:likehoodTest}).

\small
\begin{equation}
\Lambda(x, \theta) = \dfrac{L_{l}(\Theta_l^x,\Theta_l^{\theta}|x, \theta)} {L_{r} (\Theta_r^x,\Theta_r^{\theta}|x,\theta)} =  \dfrac{p_{l}(x, \theta|\Theta_l^x,\Theta_l^{\theta})} {p_{r} (x,\theta|\Theta_r^x,\Theta_r^{\theta})}, \label{eq:likehoodTest}
\end{equation}
\normalsize

Relying only on the likelihood ratio, could lead to cases where two hand-like segments are assigned the same label (left or right). To avoid this cases, and given that a frame cannot have two left nor two right hands, we follow a competitive rule in the following way. Let's assume two hands-like segments in the frame described by $z_1 = (x_1,\theta_1)$ and $z_2 = (x_2,\theta_2)$ as explained in Figure \ref{fig:geometricProblem}, and their respective likelihood ratios given by $\Lambda(x_1,\theta_1)$ and $\Lambda(x_2,\theta_2)$. The competitive ids are assigned by equation (\ref{eq:competitiveId}).

\small
\begin{eqnarray}
id_{z_1}, id_{z_2} &=& \begin{cases}
        \Lambda(x_1,\theta_1) > \Lambda(x_2,\theta_2) \rightarrow & id_{z_1} = l \\
                               & id_{z_2} = r \\ \\
        \Lambda(x_1,\theta_1) \le \Lambda(x_2,\theta_2) \rightarrow & id_{z_1} = r \\
                               & id_{z_2} = l
        \end{cases}\label{eq:competitiveId}
\end{eqnarray}
\normalsize

\section{Results}\label{sec:results}

This section evaluates our approach in two stpdf. Section \ref{sec:perfectSegmentationResults} uses the Kitchen manual masks as a perfect hand-segmenter to assess the L/R hand-identification models and the occlusion detector. In section \ref{sec:noPerfectSegmentationResults} the multi-model hand-segmenter is tuned, evaluated, and used for a realistic performance analysis of the overall system. 

\subsection{Assuming a perfect hand-segmenter} \label{sec:perfectSegmentationResults}

In this section, the extended L/R manual masks of the kitchen dataset is used as a perfect hand-segmenter. Each hand-segment is endowed with its best fitting ellipse and used as input for the L/R hand-identification model presented in section \ref{sec:LRmodel}. Table \ref{tab:maskidentification} shows the results of the L/R identification without and with likelihood ratio competition.

\begin{table}[h!]
\scriptsize
\centering
    \caption{Left and right hand identification at contour level}\label{tab:maskidentification}
\begin{tabular}{rrr|rr}
\toprule
        & \multicolumn{2}{c|}{No-Competition} & \multicolumn{2}{c}{With Competition} \\ 
\midrule
        & Left    &   Right   & Left      & Right \\
\midrule
Left    & 0.994   &   0.006   &   0.997   &   0.003\\
Right   & 0.012   &   0.988   &   0.000   &   1.000\\
\bottomrule
\end{tabular}%
\end{table}

The comparison without likelihood ratio, left side of the table, refers to the hand-identification based only on the best model. However, it is intuitive to assume that in presence of two relevant hand-like segments, they cannot be both left or right. This restriction is included by using the likelihood ratio test introduced in section \ref{sec:likehoodTest}, and presented in the right half of the table. This scheme allow us to identify almost perfectly all the masks in the dataset (i.e. $99.7\%$ of the left hands and $100\%$ of the right hands). The values reported in the table refer to the identification problem (left/right) and does not constitute a hand-segmentation.

\subsection{Without perfect segmentation}\label{sec:noPerfectSegmentationResults}

The assumption of a perfect hand-segmenter is not a realistic. Furthermore, hand segmentation is considered one of the most challenging objectives of FPV video analysis. To perform a more realistic evaluation of our approach we initially tune and evaluate the proposed $LAB$ based multi-model hand-segmenter (section \ref{sec:handSegmentationResults}). Subsequently, in section \ref{sec:occlusionDissambiguationResults} the occlusion detector is assessed to conclude with an overall evaluation of the system including each of its components.

\subsubsection{Hand-Segmentation}\label{sec:handSegmentationResults}

As presented in figure \ref{fig:multimodel} the proposed hand-segmenter is intended to alleviate the illumination problems and consequently improve the quality of the segmentation. However, some important aspects of this approach must be defined first: 

\begin{enumerate}[label=\roman*)]
    
    \item How many illumination models  ($n$) must be considered?. 
    
    \item How many models ($K$) must be provided by the KNN recommender component? 
    
    \item Which is the effect of these parameters to the quality of the segmentation?
    
\end{enumerate}

\begin{table*}
\scriptsize
\centering
    \caption{F1 Score when using different training videos.} \label{tab:f1trainingtesting}
\begin{tabular}{rrrrrrr|r}
\toprule
     &      & \multicolumn{1}{c}{\begin{sideways}Coffe\end{sideways}} & \multicolumn{1}{c}{\begin{sideways}CofHoney\end{sideways}} & \multicolumn{1}{c}{\begin{sideways}Hotdog\end{sideways}} & \multicolumn{1}{c}{\begin{sideways}Tea\end{sideways}} & \multicolumn{1}{c|}{\begin{sideways}Paelate\end{sideways}} & \multicolumn{1}{c}{\begin{sideways}Testing-F1\end{sideways}}\\ \midrule
\multicolumn{1}{c}{\multirow{5}{*}{\begin{sideways}Training\end{sideways}}} 
                     & \multicolumn{1}{c}{Coffe}    &  0.937 $\pm{0.003}$    &    0.921 $\pm{0.006}$    &    0.920 $\pm{0.002}$    &    0.941 $\pm{0.004}$    &    0.892 $\pm{0.009}$    &    \textbf{0.920 $\pm{0.005}$}\\
\multicolumn{1}{c}{} & \multicolumn{1}{c}{CofHoney} &  0.933 $\pm{0.004}$    &    0.925 $\pm{0.005}$    &    0.917 $\pm{0.002}$    &    0.931 $\pm{0.004}$    &    0.864 $\pm{0.009}$    &    0.914 $\pm{0.003}$\\
\multicolumn{1}{c}{} & \multicolumn{1}{c}{Hotdog}   &  0.923 $\pm{0.004}$    &    0.910 $\pm{0.006}$    &    0.930 $\pm{0.002}$    &    0.925 $\pm{0.006}$    &    0.883 $\pm{0.011}$    &    0.912 $\pm{0.005}$\\
\multicolumn{1}{c}{} & \multicolumn{1}{c}{Tea}      &  0.925 $\pm{0.009}$    &    0.909 $\pm{0.008}$    &    0.910 $\pm{0.003}$    &    0.935 $\pm{0.005}$    &    0.899 $\pm{0.006}$    &    0.911 $\pm{0.005}$\\
\multicolumn{1}{c}{} & \multicolumn{1}{c}{Pealate}  &  0.918 $\pm{0.008}$    &    0.906 $\pm{0.009}$    &    0.902 $\pm{0.011}$    &    0.922 $\pm{0.008}$    &    0.904 $\pm{0.005}$    &    0.913 $\pm{0.007}$\\
\bottomrule
\end{tabular}%

\end{table*}

In order to answer these questions a computational experiment is designed to tune $n$ and $K$ and evaluate proposed multi-model approach with the state-of-the-art hand-segmentation methods. For the experiment, the subject $1$ of the kitchen dataset is used. We train a multi-model hand segmenter using each video for training and the remaining ones for testing. As explained in section \ref{sec:handSegmentation} each illumination model is a random forest, which introduces a random component to the hand-segmenter. To alleviate the randomness in the evaluation, the training-testing is executed $5$ times with different random seeds. With this in mind, given ($n$) and ($K$), a total of $25$ training and testing errors are obtained (e.g. $5$ per training video times $5$ per random seed).

\begin{figure}[h!]
    \centering
    \includegraphics[width=1\linewidth]{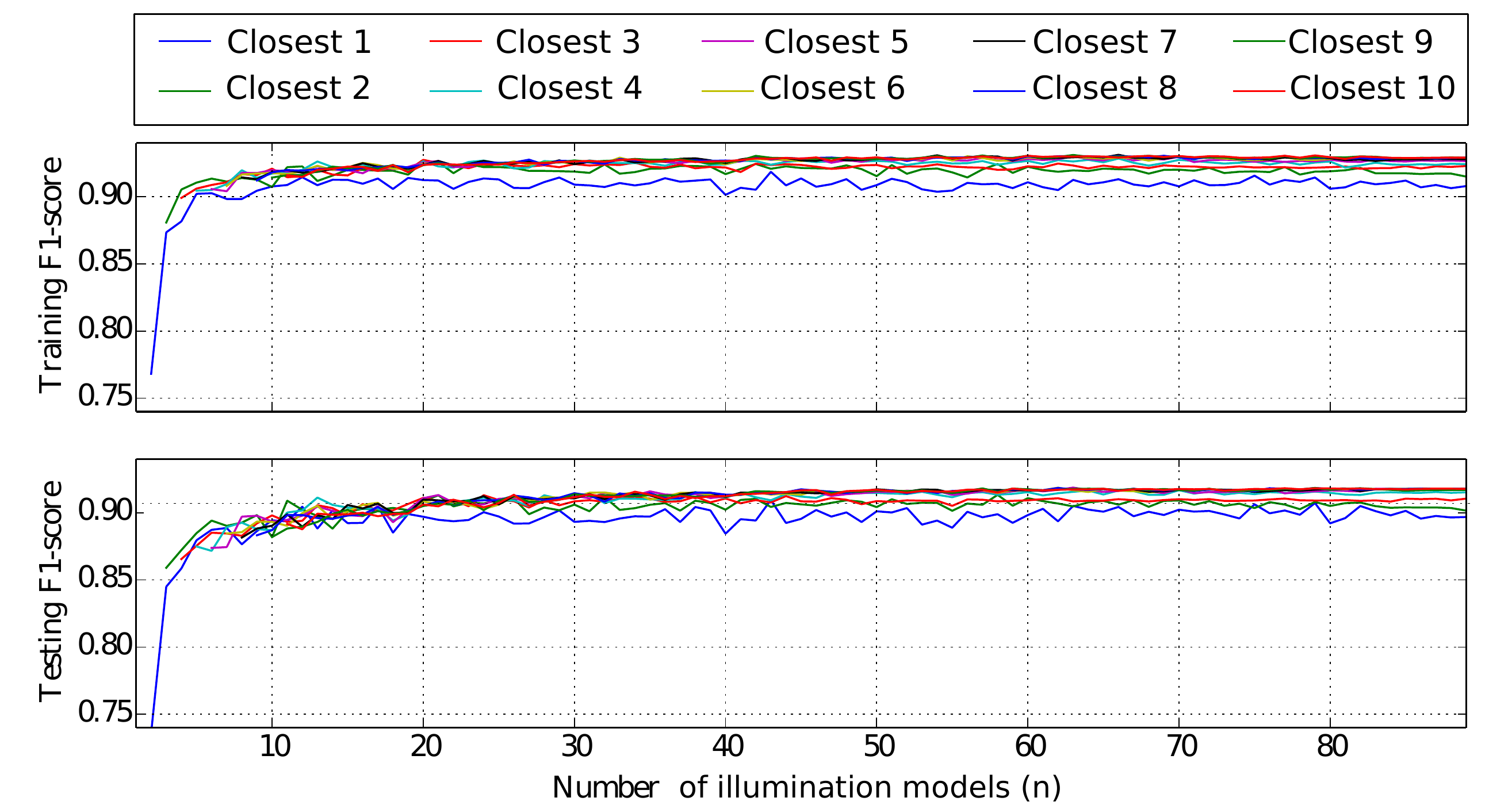}
    \caption{Hand-segmentation $F1$ score when changing the number of illumination models ($n$) and the number of closest models ($K$) to fuse. The first and second plots are the training and testing $F1$ scores, respectively. The number of illumination models $n$ is plotted in the horizontal axis, while the $F1$ score is in the vertical axis; the colors represent the number of models to fuse $K$.} \label{fig:handSegmenterVsModels}
\vspace{-10pt}
\end{figure}

Figure \ref{fig:handSegmenterVsModels} shows the average training (top plot) and testing (bottom plot) $F1$ scores while changing the number of illumination models ($n$). The colors of the lines (legend of the figure) refer to the use of the $K$ closest random forests in the fusion part. The image shows a quick improvement in the performance when the number of illumination models increases. As reference, the testing error changes from $75\%$ to $92.8\%$ when using $20$ illumination models instead of a single one. For the remaining part of this paper, the number of illumination models is set to $20$. Regarding the number of illumination models to fuse $K$, the performance quickly converges on $K=5$; concluding that, for the kitchen dataset, the fusion of more than $5$ illumination models does not provide additional improvements to the segmentation quality. In total the fusion of $5$ illumination models contributes two units in the $F1$ score compared with the use of only $1$. In the remainder of this paper a value of $K=5$ is used.

For what concerns the training video selection, Table \ref{tab:f1trainingtesting} shows a detailed comparison of the binary segmentation performances when trained with different videos. The table shows the mean $F1$ and its standard deviation. The diagonal of the table is training $F1$ while, and the remaining values are the testing $F1$ scores. The overall testing $F1$ is in the final column. Results allow us to conclude that the choice of the testing video does not create a substantial effect on the overall performance. The latter is true if the light conditions of the videos are similar. In the remainder of the paper, we use the ``Coffee'' video sequence ($4700$ frames - $79$ masks - $7$ occluded masks) as training sequences, and the remaining ``CofHoney'', ``Hotdog'', ``Tea'', ``Pealate'' for testing. The testing sequences contain in total $21240$ frames, $367$ Left/Right masks and $44$ occluded masks. Please refer to \cite{Fathi2012a} for extra details about the Kitchen dataset.


Finally, Table \ref{tab:comparison} compares the multi-model hand-segmenter with previous works. If compared with the single pixel-by-pixel classifier of \cite{Li2013a}, our approach achieves improvements between $3$ and $5$ $F1$ score points. After the post-processing, our method achieves a total improvement of $9$, $12$ and $14$ $F1$ points on the ``Coffee'', ``Tea'' and ``Peanut'' video sequences, respectively.  In comparison to the shape aware hand-segmenter proposed by \cite{Zhu2014}, our implementation performs better in all the video sequences. In particular, the ``Tea'' video sequence is improved by $10$ $F1$ points.

\begin{table}
\scriptsize
\centering
    \caption{Hand-Segmenter state of the art comparison. The performances reported for the state-of-the-art are taken from \cite{Zhu2014}} \label{tab:comparison}
\begin{tabular}{llccc}
\toprule
                                                                 &  Coffee           &  Tea             &  Peanut                     \\ \midrule
 1999 - Single pixel color \cite{Jones1999}                      &   0.83            &  0.80            &   0.73                      \\
 2011 - stabilization + gPb + superpixel + CRF \cite{Fathi2011a} &   0.71            &  0.82            &   0.72                      \\
 2013 - Li $1\times1$ window \cite{Li2013a}                      &   0.85            &  0.82            &   0.74                      \\
 2013 - Li $9\times9$ window \cite{Li2013a}                      &   0.88            &  0.88            &   0.76                      \\
 2014 - Shape Aware Forest (post-process)  \cite{Zhu2014}        &   0.90            &  0.84            &   0.84                      \\ \midrule
 2016 - Ours (k=20, m=50)                                        &   0.88            &  0.87            &    0.77             \\
 2016 - Ours (k=20, m=50) + Hand-Id Post Process                 &   \textbf{0.94}   &  \textbf{0.94}   &  \textbf{ 0.88  }             \\
\bottomrule
\end{tabular}%
\end{table}

\subsubsection{Occlusions and overall performance} \label{sec:occlusionDissambiguationResults}

The extended masks can be used to identify evaluation cases for the occlusion detector and the splitting method. To evaluate the occlusion detector we initially select the masks with hand-to-hand occlusions and check if the occlusion detector finds them.  In total, the ``Kitchen'' dataset (subject $1$) contains $51$ hand-to-hand occluded frames, and the algorithm \ref{alg:occlusion} identifies $98\%$.

When automatically segmented, the silhouette of the hands will be affected by the false-positives and false-negatives, and as the consequence, some extra frames could be mis-detected as hand-to-hand occlusions (i.e., two noise hand-like segments close enough to be considered occluded). This is not a problem, since the algorithm will split these cases as a real occlusions and only some extra computational time is needed.

\begin{table}[h!]
\scriptsize
\centering
\caption{Evaluation of the hand segmentation only when split is required} \label{tab:occlusion}
\begin{tabular}{r|rrr}
\toprule
        & No-Hand & Left      & Right   \\
\midrule
No-Hand & 0.984   &   0.007   &   0.009 \\
Left    & 0.058   &   0.934   &   0.009 \\
Right   & 0.080   &   0.006   &   0.914 \\
\bottomrule
\end{tabular}%
\end{table}

\begin{table}[h!]
\scriptsize
\centering
    \caption{Effect of the occlusion detection and dissambiguation in the overall performance. } \label{tab:disambiguation}
\begin{tabular}{r|rrr|rrr}
\toprule
     & \multicolumn{3}{c|}{Without split} & \multicolumn{3}{c}{With split} \\
\midrule
        & No-Hand & Left & Right & No-Hand & Left & Right \\
\midrule
No-Hand & 0.992   &   0.004   &   0.004   &   0.992   &   0.004   &   0.004 \\
Left    & 0.073   &   0.821   &   0.106   &   0.073   &   0.923   &   0.004 \\
Right   & 0.096   &   0.066   &   0.838   &   0.096   &   0.001   &   0.903 \\
\bottomrule
\end{tabular}%
\end{table}

To evaluate the hand-to-hand occlusion split the extended L/R masks was used as ground truth to perform $3$ class pixel-by-pixel classification analysis (i.e., background, left hand, right hand). First, in seek of a better evaluation of the split procedure, only the frames detected as occluded are used. Table \ref{tab:occlusion} shows the confusion matrix of the $3$ class pixel-by-pixel segmentation for all the frames detected as occlusion.  The table concludes that, in the case of occlusion, the split leads to a proper classification of $93.4\%$ and $91.4\%$ of the left-hand and right-hand pixels, respectively. It is important to note, as shown previously, that the main cause of the misclassified left/right pixels is not the split procedure, but the noisy segmentation. 

\begin{table*}
\scriptsize
\centering
    \caption{L/R hand-segmentation confusion matrix. This table uses the ``Coffe'' video sequence for training} \label{tab:finalsegm}
\begin{tabular}{rr|rrr|rrr|rrr|rrr|rrr}
\toprule
   &  & \multicolumn{3}{c|}{CofHoney} & \multicolumn{3}{c|}{Hotdog} & \multicolumn{3}{c|}{Tea} & \multicolumn{3}{c|}{Pealette} & \multicolumn{3}{c}{Total} \\
\midrule
   &  & \multicolumn{1}{c}{\begin{sideways}No-hands\end{sideways}} & \multicolumn{1}{c}{\begin{sideways}Left\end{sideways}} & \multicolumn{1}{c|}{\begin{sideways}Right\end{sideways}} & \multicolumn{1}{c}{\begin{sideways}No-hands\end{sideways}} & \multicolumn{1}{c}{\begin{sideways}Left\end{sideways}} & \multicolumn{1}{c|}{\begin{sideways}Right\end{sideways}} & \multicolumn{1}{c}{\begin{sideways}No-hands\end{sideways}} & \multicolumn{1}{c}{\begin{sideways}Left\end{sideways}} & \multicolumn{1}{c|}{\begin{sideways}Right\end{sideways}} & \multicolumn{1}{c}{\begin{sideways}No-hands\end{sideways}} & \multicolumn{1}{c}{\begin{sideways}Left\end{sideways}} & \multicolumn{1}{c|}{\begin{sideways}Right\end{sideways}} & \multicolumn{1}{c}{\begin{sideways}No-hands\end{sideways}} & \multicolumn{1}{c}{\begin{sideways}Left\end{sideways}} & \multicolumn{1}{c}{\begin{sideways}Right\end{sideways}} \\ \midrule
\multirow{3}{*}{\begin{sideways}60 FPS\end{sideways}}
& No-hands & 0.990   &   0.003   &   0.007   &   0.989   &   0.005   &   0.006   &   0.996   &   0.002   &   0.002   &   0.991   &   0.006   &   0.003   &   \textbf{0.992}   &   \textbf{0.004}   &   \textbf{0.004}\\
& Left     & 0.064   &   0.932   &   0.004   &   0.040   &   0.958   &   0.002   &   0.056   &   0.943   &   0.001   &   0.120   &   0.871   &   0.009   &   \textbf{0.073}   &   \textbf{0.923}   &   \textbf{0.004}\\
& Right    & 0.092   &   0.002   &   0.906   &   0.136   &   0.001   &   0.864   &   0.082   &   0.000   &   0.918   &   0.112   &   0.002   &   0.886   &   \textbf{0.096}   &   \textbf{0.001}   &   \textbf{0.903}\\
\midrule
\multirow{3}{*}{\begin{sideways}30 FPS\end{sideways}}
& No-hands & 0.990   &   0.003   &   0.007   &   0.989   &   0.006   &   0.005   &   0.996   &   0.002   &   0.002   &   0.991   &   0.006   &   0.003   &   0.992   &   0.004   &   0.004 \\
& Left     & 0.064   &   0.930   &   0.006   &   0.039   &   0.958   &   0.002   &   0.057   &   0.932   &   0.011   &   0.119   &   0.874   &   0.007   &   0.073   &   0.921   &   0.007 \\
& Right    & 0.093   &   0.009   &   0.898   &   0.133   &   0.002   &   0.865   &   0.082   &   0.000   &   0.918   &   0.109   &   0.003   &   0.888   &   0.095   &   0.004   &   0.900 \\
\midrule
\multirow{3}{*}{\begin{sideways}15 FPS\end{sideways}}
& No-hands & 0.990   &   0.003   &   0.007   &   0.990   &   0.006   &   0.005   &   0.996   &   0.002   &   0.002   &   0.991   &   0.006   &   0.003   &   0.993   &   0.004   &   0.004\\
& Left     & 0.063   &   0.919   &   0.017   &   0.040   &   0.914   &   0.047   &   0.056   &   0.940   &   0.004   &   0.118   &   0.865   &   0.017   &   0.072   &   0.907   &   0.021\\
& Right    & 0.092   &   0.008   &   0.900   &   0.140   &   0.057   &   0.803   &   0.081   &   0.000   &   0.919   &   0.109   &   0.036   &   0.855   &   0.096   &   0.015   &   0.889\\
\bottomrule
\end{tabular}%
\end{table*}

To conclude, Table \ref{tab:disambiguation} shows the benefit of the occlusion detection and the split to the overall hand-identification. The first vertical group ignores the occlusion problem, while the second is obtained using proposed approach. Both confusion matrix are identical in the background performance since the hand-segmenter is the same for the two experiments. The L/R hand-segmentation gains almost ten percentage points when occlusions are considered. Eventually, table \ref{tab:finalsegm} provides the detailed results for each testing video. For comparative purposes, the table provides the performances obtained by using $60$, $30$ and $15$ frames per second. It can be noticed that the overall performance is not considerably affected by a sampling rate of $30fps$. When using $15$ frames per second, the segmentation quality suffers a small reduction, but the throughput of the system is considerably improved. All the results reported in this paper use a latency of $60fps$ (bold digits).


\section{Conclusions and future research}\label{sec:conclusions}

This work presented a hierarchical strategy to segment and identify the left and right hands of the user in egocentric videos. The proposed method provides valuable information about the hand-usage and opens the door to use wearable cameras in applications involving bi-manual tasks, for example for driving applications or medical therapy for upper limb mobility problems.

The first level of proposed method is a multi-model structure that delineates the hand-like pixels on each egocentric frame. Experimental results show that proposed multi-model implementation, jointly with the hand-identification post-processing, achieves $F1$ scores of around $0.92$, which constitutes a significant improvement to the shape-aware classifier proposed in \cite{Zhu2014}.

The second level, executed if required, is the hand-to-hand occlusion identification and disambiguation. The experimental section shows the importance of this step to understand the hands of the user as two cooperative entities working jointly to accomplish a particular task. Our results indicate that, by handling hand-to-hand occlusions, it is possible to obtain improvements around $10\%$ in L/R hand-segmentation.

The final level, the hand-identification, relies on a Maxwell function of angle and horizontal position, to decide whether a hand-like segment is left or right. Experimental results show that our L/R identification model identifies with $99\%$ certainty if a hand is left or right. We highlight this as a considerable improvement regarding efficiency and accuracy to the state of the art, where a SVM is used to understand the state of the hands as: i) only left, ii) only right, iii) both hands.


As a future research line, we highlight the use of the identified hands as cooperative entities to understand how the user is performing a particular task. The results obtained with our method can be used as the measurement model in the framework of tracking interacting objects to get reliable hands trajectories and augmented states. These trajectories could lead to a proper understanding of the user's hands movements, which constitutes a starting point to use wearable cameras in medical therapy. Based on our current research the hand-tracking level requires considerable development in the definition of the dynamic models ruling the non-linear movement of the hands. Additional issues must be solved when noisy measurements are detected or in the presence of complex hand interactions.

Finally, some of the methods presented in this paper, such as the multi-model classification algorithm, could be applied in more general scenarios. The objective of this paper is to exploit the advantageous location of the camera to extract additional information about the hands of the user. The use of the segmentation model in other video perspective or application is left as an interesting future work.

\section{Acknowledgement}

This work was partially supported by the Erasmus Mundus joint Doctorate in
Interactive and Cognitive Environments, which is funded by the EACEA, Agency of
the European Commission under EMJD ICE.

The authors thank the Cyberinfrastructure Service for High Performance
Computing, ``Apolo'', at EAFIT University, for allowing us to run our
computational experiments in their computing centre.



\section*{Bibliography}
\footnotesize
\bibliographystyle{IEEEbib}
\bibliography{Betancourt2016}

\end{document}